# Chemistry in Protoplanetary Disks

Ke Zhang

University of Wisconsin-Madison
475 N Charter St, Madison, WI 53706 USA ke.zhang@wisc.edu

Abstract: Planets are formed inside disks around young stars. The gas, dust, and ice in these natal disks are the building materials of planets, and therefore their compositions fundamentally shape the final chemical compositions of planets. In this review, we summarize current observations of molecular lines in protoplanetary disks, from near-infrared to millimeter wavelengths. We discuss the basic types of chemical reactions in disks and the current development of chemical modeling. In particular, we highlight the progress made in understanding snowline locations, abundances of main carriers of carbon, oxygen, and nitrogen, and complex organic molecules in disks. Finally, we discuss efforts to trace planet formation history by combining the understanding of disk chemistry and planet formation processes.

## Overview: Chemistry in Planet-forming disks

The habitability of a planet depends not only on its physical properties such as mass, radius, and orbital location, but also on its chemical properties, such as its water and organic contents on its surface and atmosphere. A fundamental question of the planet formation field revolves around how planet formation processes determine the chemical compositions of nascent planets. Planets are formed inside disks around young stars (see reviews by Williams and Cieza, 2011; Andrews, 2020), the so-called protoplanetary disks. In these natal disks, the gas, dust, and ice serve as the raw building materials of planets. The compositions of these raw materials are first inherited from the compositions of the host stars themselves (see more details in the chapter "Host Stars and How Their Compositions Influence Explanets" by Hinkel et al.). The materials are then further altered and redistributed at the stage of protoplanetary disks (e.g., Krijt et al., 2023; Oberg et al., 2023).¨ Therefore, a thorough understanding of chemistry in these natal disks is an essential part of our understanding of planet formation and the potential habitability of extrasolar planetary systems.

The field of chemistry in protoplanetary disks has two goals. The first one is to understand the compositions of materials in natal disks, and how the compositions of disk materials change over location and time in disks during planet formation. The second one, which goes beyond the pure chemistry aspect, aims to understand how we can use molecular and atomic lines as probes to study physical conditions and processes during planet formation, which are otherwise hard or even impossible to constrain. The two goals are intertwined as physical and chemical processes are often coupled with each other.

In this review, we discuss recent observations of chemistry in protoplanetary disks, theoretical expectations of chemical reactions in protoplanetary disks, the latest important insights into chemistry in disks, and connections between disk chemistry to compositions of exoplanets and solar system objects. Finally, we highlight a few promising frontiers in the field for the upcoming decade. There have been excellent recent reviews on chemistry in protoplanetary disks (e.g., Henning and Semenov, 2013; Oberg and Bergin, 2021; Oberg et al., 2023). We are not trying to replicate these efforts; instead, this review aims to provide an easily accessible introduction material for interdisciplinary readers.

## Basics of protoplanetary disks

Planets are born in gas- and dust-rich disks around young stars, the so-called protoplanetary disks (see reviews by Williams and Cieza, 2011; Andrews, 2020). These disks are analogs of our Solar system at 4.6 billion years ago, providing precious windows for us to witness the formation of planets. The gas, dust, and ice in these disks are raw materials to build planets. As a disk evolves, the amount of disk materials and their compositions may change dramatically. As a result, the final internal structure/atmosphere of a planet depends on how, when, and what types of materials are accreted.

The chemical compositions and structures of protoplanetary disks are largely determined by the physical properties of the disk, in particular the structures of density, temperature, and radiation field (e.g., Bergin et al., 2007). The characterization of disk physical properties, such as the gas and dust mass distribution, has made revolutionary progress over the past decade, thanks to the Atacama Large Millimeter/submillimeter Array (ALMA) (see review by Miotello et al., 2023). So far, ~1500 disks (in the nearby star-forming regions within 500pc) have been observed by ALMA at millimeter continuum, and around a few hundred disks with at least CO molecular line observations (e.g., Ansdell et al., 2016; Barenfeld et al., 2016; Long et al., 2017; Cieza et al., 2019; Tobin et al., 2020). A large diversity of physical properties have been seen among protoplanetary disks. For example, even in the same stellar mass, there are two orders of magnitude differences in the dust disk masses. This diversity in physical properties would naturally lead to diversity in chemical structures, which contributes to the large diversity seen in extrasolar planets.

The gas in protoplanetary disks can last for <~10Myr (e.g., Fedele et al., 2010; Ercolano and Pascucci, 2017). Gas can be accreted onto the central star, be used to form gas envelop of planets, and be blown away by disk winds. The characteristic lifetime scale of protoplanetary disks is estimated to be ~3Myr, based on the decreasing fraction of young stars with near-Infrared excess or stars showing accretions. This lifetime scale sets a fundamental limit for the timescale of the formation of giant planets. The amount of dust masses (grains with sizes up to 1cm) is found to generally decrease over time, with a median dust mass of $1M_\oplus$ for 1-3Myr disks to a median of $0.1M_\oplus$ for 5-10Myr disks (Manara et al., 2023). The gas masses of protoplanetary disks are still highly uncertain, because the dominant mass constitute $H_2$ does not have a strong emission line at the bulk temperature of disks (e.g., Bergin and Williams, 2017; Trapman et al., 2022). Most of the gas disk estimations are based on dust mass and the assumption of a gas-to-dust mass of 100, which are probably upper limits. The majority of disks are likely to have a lower gas mass compared to the Minimum Solar Nebular Mass (~0.01$M_\odot$, the minimum mass required to form solar system plants, Hayashi 1981).

Besides the large intrinsic spread in global disk properties, there are other complications in the study of chemistry in protoplanetary disks. One complication is that these disks have large gradients in the temperature, density, and radiation fields, in the radial as well as the vertical direction (e.g., Calahan et al., 2021; Law et al., 2021). For example, the inner 1au region can be quite hot, reaching 500-1500K (e.g., Salyk et al., 2008; Carr and Najita, 2008; Dullemond and Monnier, 2010), while the bulk region outside 100au is below 20K (e.g., Dartois et al., 2003; Pinte et al., 2018; Law et al., 2021). As a result, the chemical compositions of a given disk vary with radial and vertical locations. Another complication is that the physical conditions in the disk evolve over time at a comparable timescale as many chemical reactions. As a result, our predictive power on chemistry in disks also depends on our understanding of the physical processes in disks.

## Observations of molecules in protoplanetary disks

Unlike studies of many objects in the Solar System objects, where in-situ samples can be collected and measured, the observations of protoplanetary disks are solely based on remote sensing. Molecules in protoplanetary disks have been detected at a wide range of wavelengths, from UV to cm wavelengths (see review by Henning and Semenov, 2013). However, lines within a given wavelength range, mid-IR, only trace a limited range of radial and vertical locations in the disk, due to large temperature and density gradients in protoplanetary disks. For example, the mid-IR traces lines mostly arise from warm gas (T>500K), which is found in the inner few au around T Tauri stars, while the pure rotational lines of heavy molecules are found from the colder and outer regions of disks. As a result, our observations of protoplanetary disks are similar to blind men's observations of an elephant, and therefore multi-wavelength observations are needed to characterize the chemical structures at different regions of protoplanetary disks.

Infrared and millimeter wavelengths are the two primary windows to probe chemistry in protoplanetary disks. The vibrational bands of molecular lines are mainly at infrared wavelengths and pure rotational lines at millimeter wavelengths. Below we describe the main molecules detected within these wavelength ranges.

# Infrared observations of protoplanetary disks near-IR

wavelength range (1-5$\mu m$)

At the near-IR wavelength range (1-5$\mu m$), CO is the most commonly detected molecular line, seen in many protoplanetary disks around 0.5-3$M_\odot$ stars (e.g., Carr et al., 1993; Blake and Boogert, 2004; Brittain et al., 2007; Salyk et al., 2011; Brown et al., 2013; Banzatti and Pontoppidan, 2015). The fundamental band of rovibrational lines ($v$=1-0) at 4.7$\mu m$ is the most frequently detected feature, and the overtone band ($v$=2-0) at 2.3$\mu m$ is also detected in a small number of protoplanetary disks. The CO line profiles can be spectrally resolved by high-resolution ground-based observations (R~25,000-100,000), which provides additional information to study the velocity field of the CO-emitting areas. The line shapes of CO rovibrational lines often show two distinctive components, a narrow component with FWHM of 10-50km/s, consistent with Keplerian rotation between 0.1- a few au, and a broad velocity component with FWHM of 50-200km/s, arising from ~0.05au (e.g., Banzatti et al., 2022).

$H_2O$ has three vibrational modes, and the symmetric and asymmetric stretching bands between 2.5-3.5$\mu m$ have been detected in many disks from ground-based observations (e.g., Salyk et al., 2008, 2019; Banzatti et al., 2022). These detections suggest that hot water vapor (~1000K) commonly exists inside 0.5au region of protoplanetary disks. The detections of organic molecules at near-IR wavelengths have been challenging from ground-based observations. HCN and $C_2H_2$ were detected only in a few disks (Mandell et al., 2012). $CH_4$ has only been detected in absorption features in one disk (Gibb and Horne, 2013).

mid-IR wavelength (5-30$\mu m$)

At mid-infrared wavelength range (5-30$\mu m$), simple molecules like $H_2O$, CO, $CO_2$, HCN, $C_2H_2$, OH, and $H_2$, have been observed in ~ 100 protoplanetary disks by the Spitzer space telescope with a spectral resolution of R~600 (e.g., Carr and Najita, 2008; Salyk et al., 2008; Pontoppidan et al., 2010; Carr and Najita, 2011; Salyk et al., 2011). These molecules are detected via their rovibrational or rotational lines at mid-IR. The upper energy levels of these lines are generally very high, spanning over 500-10,000K (Pontoppidan et al., 2010). Therefore, these lines are expected to be emitted from the hot region inside a few aus from the central star. Simple slab models (assuming constant temperature and density) showed that typical excitation temperatures between 400-1600K and emitting areas with a radius around 1au (e.g., Salyk et al., 2011; Carr and Najita, 2011). In a few cases, a small number of mid-IR water lines are also observed by the ground-based highresolution spectrograph, which provided direct constraints on the line profiles and therefore the line emitting regions (Pontoppidan et al., 2011; Salyk et al., 2019). These lines showed broad line profiles with FWHM~ 20-35km/s, consistent with the velocity range of Keplerian rotation from the inner 1au region. In short, the mid-IR lines provide an important window to trace chemical compositions in the terrestrial planet-forming regions.

Interestingly, Spitzer's observations showed that detection rates of molecular lines in disks are correlated with the masses of the host stars. The detection rates of $H_2O$ and small organic are generally high (∼30-60%) in T Tauri disks (stellar mass below $1M_\odot$), but drop to nearly zero for Herbig disks (stellar mass between $2-3M_\odot$) (Pontoppidan et al., 2010). It is still under debate whether the dearth of water detection in Herbig disks is due to intrinsic difference in the water abundance between the two types of disks or due to the high IR continuum in the more luminous Herbig disks blanketing the weak molecular lines (e.g., Antonellini et al., 2016). Another interesting dependence of stellar masses was the $HCN/H_2O$ line ratios seen in disks around late M-dwarf stars ($0.3M_\odot$ or smaller) on average are significantly higher than that of the solar-like type stars, suggesting a higher C/O elemental ratio in the gas of inner disk regions in late M-dwarf sources (Pascucci et al., 2013). However, the late M-dwarf sample size was small (∼10) and more observations are needed to confirm the trend.

The inventory of the molecules detected in mid-IR is expected to significantly expand over the next few years, thanks to the newly launched James Webb Space Telescope (JWST). Compared to Spitzer, JWST offers an order of magnitude higher sensitivity and a factor of a few higher spectral resolutions. See Figure 1 for an example of the JWST/MIRI spectrum of a protoplanetary disk.

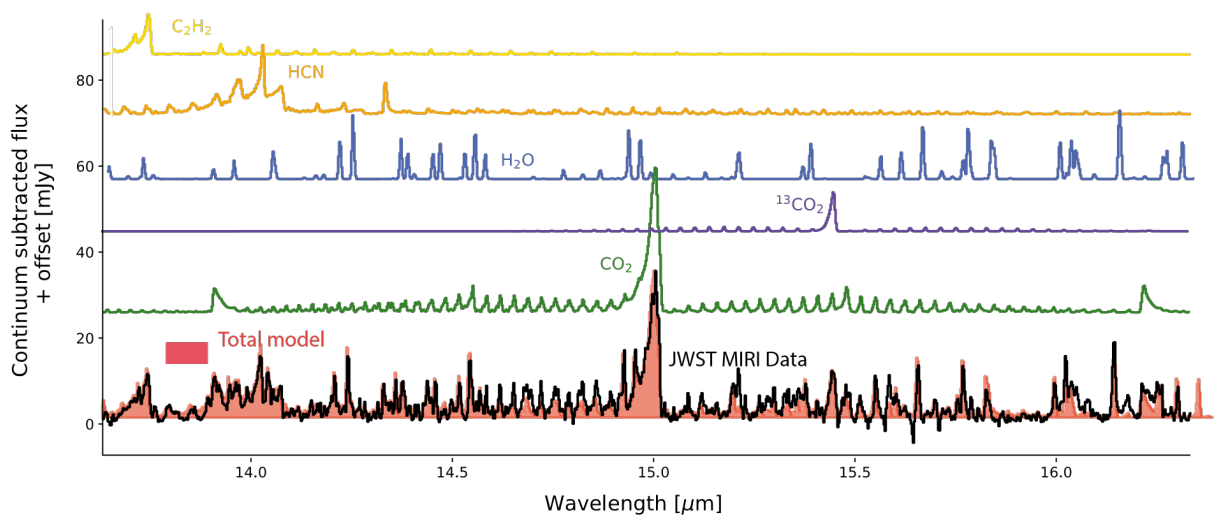

Figure 1: *JWST MIRI/MRS continuum-subtracted spectrum of the GW Lup disk between 13.5–16.5µm (Grant et al., 2023). The GW Lup protoplanetary is located in the Lupus region (1-3Myr). The JWST-MIRI data (black) is compared to a slab model (red) composed of different molecules.*

The recent JWST discoveries include the first detection of $^{13}CO_2$ in a disk around a solar-like star (Grant et al., 2023), and a suite of organics were detected in a disk around a very low mass young star (M = $0.14M_\odot$), including $C_4H_2$, $C_6H_6$, $^{13}C^{12}CH_2$, and a tentative detection of $CH_4$ (Tabone et al., 2023). Also, JWST recently detected $CH^+_3$ in a protoplanetary disk in the Orion star-forming region, supporting that gas-phase organic chemistry is activated by ultraviolet radiation (Berné et al., 2023).

Far-IR wavelength range (30-550$\mu m$)

In the far-IR wavelength range (30-550$\mu m$), most of the disk observations were provided by the Herschel Space Telescope. The HD $J$= 1-0 line was detected in three protoplanetary disks, which provided important constraints on the gas disk masses (Bergin et al., 2013; McClure et al., 2016). The far-IR wavelength range covers a wide range of water lines ($E_{up}$ = 53-1100K) that trace cold water vapors (150K or lower) beyond the surface water snowline (Zhang et al., 2013; Blevins et al., 2016). However, there were few far-IR water line detections in protoplanetary disks, compared to the theoretical predictions. The cold water vapor was only detected in the TW Hya protoplanetary disk (a T Tauri disk), the HD 100546 disk (a Herbig disk) and tentatively in the HD 163296 disk (Herbgi disk) (Bergin et al., 2010; Hogerheijde et al., 2011; Fedele et al., 2012; Riviere-Marichalar et al., 2012; Meeus et al., 2012; Du et al., 2017). High $J$ rotational lines of CO are commonly detected in Herbig disks and in some T Tauri disks (e.g., Meeus et al., 2012; Fedele et al., 2012; Bruderer et al., 2012; Bergin et al., 2013). $NH_3$ and $N_2H^+$ were detected in the TW Hya disk (Salinas et al., 2016). The dearth of cold water vapor lines was the first evidence that volatile may be largely depleted in the atmosphere of the outer disk region. Please see more discussions in Section 4.2.

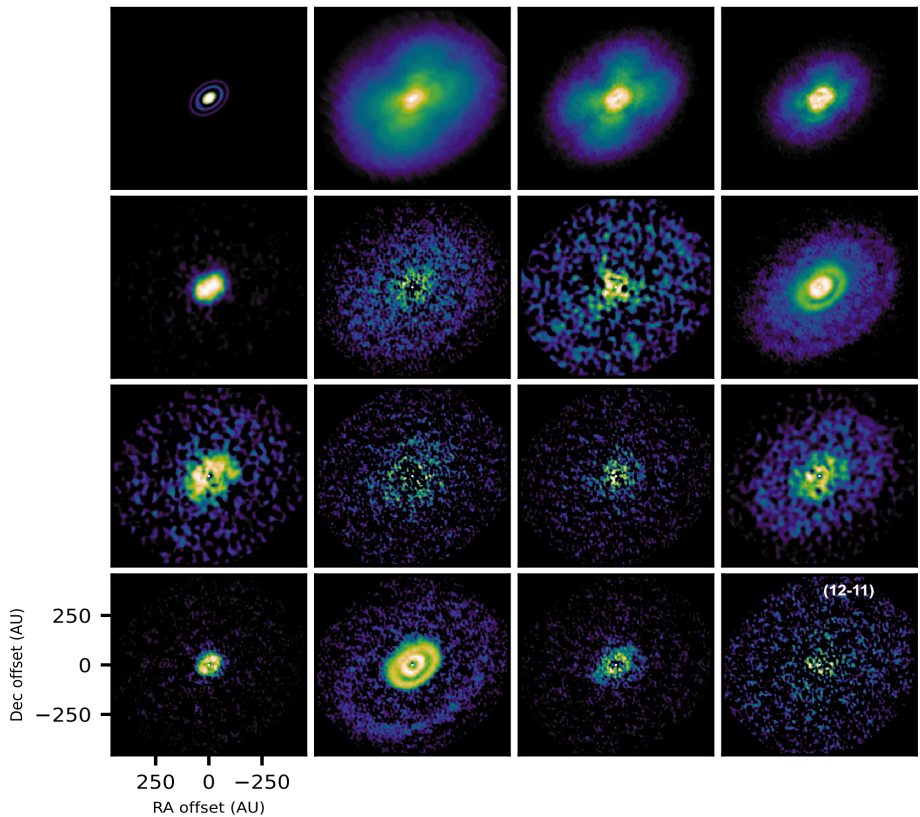

Figure 2: 15 molecular faces of the HD 163296 disk, data are adopted from Oberg et al. (2021).¨ These comprise a representative, but non-exhaustive, set of zeroth moment maps towards the HD 163296 disk. The diversity of molecular emission are due to the differences in the temperature of the emission region and abundance distributions of molecules.

## Millimeter wavelength observations of protoplanetary disks

Another important window to probe chemistry in protoplanetary disks is the millimeter wavelength range (~0.3-3mm). A great advantage of the mm-wavelength observations is that they provide spatially resolved images (down to 30milliarcsec) and high spectral resolution (down to 0.01km/s). This observational window has made revolutionary progress over the past decade, thanks to the Atacama Large Millimeter/submillimeter Array (ALMA). ALMA can observe molecular line emissions of nearby protoplanetary disks with spatial resolution down to 10-20au (e.g., Oberg et al.,¨ 2021). The spatially resolved observations provide important constraints to the chemical abundances, temperature structures, and gas mass distributions in the 20-400au regions of protoplanetary disks (e.g., Dartois et al., 2003; Schwarz et al., 2016; Pinte et al., 2018; Oberg et al., 2021;¨ Zhang et al., 2021; Law et al., 2021). Figure 2 shows examples of ALMA images of various of molecular line emissions from the HD 163296 protoplanetary disk.

In mm wavelength observations, $^{12}$CO is still the most wildly detected molecule in protoplanetary disks, ubiquitously seen in hundreds of nearby protoplanetary disks (e.g., Ansdell et al., 2016; Barenfeld et al., 2016; Long et al., 2017). Its main isotopologues, $^{13}$CO and C$^{18}$O are also commonly detected. The even more rare isotopologues, C$^{17}$O, $^{13}$C$^{18}$O, $^{13}$C$^{17}$O, have been detected in several bright protoplanetary disks (e.g., Zhang et al., 2020; Booth et al., 2019; Booth and Ilee, 2020; Zhang et al., 2021). Due to its high abundance and stable chemical nature, CO and its isotopologue lines are usually used as a probe to study the distribution of gas mass and temperature in protoplanetary disks (e.g., Dartois et al., 2003; Schwarz et al., 2016; Zhang et al., 2017; Pinte et al., 2018; Zhang et al., 2019, 2021; Law et al., 2021).

Simple polyatomic molecules are also commonly detected in protoplanetary disks, including H$_2$CO, HCN, DCN, HNC, CS, C$_2$H, c-C$_3$H$_2$, CN, DCN (e.g., Dutrey et al., 1997; Oberg et al.,¨ 2010; Cleeves et al., 2016; Bergin et al., 2016; van Terwisga et al., 2019; Pegues et al., 2020; Guzman et al., 2021). These molecules are often studied with CO isotopologue lines together, to´ measure their relative abundances to CO, and to constrain the general elemental C/O, C/H, and N/H abundances in the disk gas (e.g., Bergin et al., 2016; Cleeves et al., 2018; Miotello et al., 2019; Bosman et al., 2021).

For molecular ions, N$_2$H$^+$, HCO$^+$, DCO$^+$, and H$^{13}$CO$^+$, are the most commonly detected species in protoplanetary disks (e.g., Oberg et al., 2010; Qi et al., 2013, 2015, 2019; Huang et al.,¨ 2017; Anderson et al., 2019, 2022). These ions provide important constraints on the ionization rate in protoplanetary disks (e.g., Cleeves et al., 2015, 2017; Seifert et al., 2021; Trapman et al., 2022). Besides as an ionization tracer, N$_2$H$^+$ is also used as a tracer to probe the mid-plane CO snowline and the gas-phase abundance of CO, because its formation and destruction are closely related to gas phase CO (e.g., Qi et al., 2013, 2015, 2019; van 't Hoff et al., 2017; Anderson et al., 2019, 2021; Trapman et al., 2022).

The detections of H$_2$O in disks via ground-based millimeter observations are shown to be really challenging, especially in older than 1Myr disks (Notsu et al., 2016, 2019; Carr et al., 2018).

HDO and H$^{18}_2$O lines have been detected in a young disk (<1Myr-old) which is particularly warm due to its ongoing accretion outburst (Tobin et al., 2023).

This is of great interest to study the abundances of Complex Organic molecules (COMs), defined as unsaturated carbon-bearing molecules with five or more atoms (Herbst and van Dishoeck, 2009). However, COMs appear to be difficult to detect in disks, likely because most of the COMs freeze out on grain surfaces and therefore only have very small abundances in the gas phase. CH$_3$CN and HC$_3$N have only detected in ~10 disks (Oberg et al., 2015; Bergner et al., 2018;¨ Loomis et al., 2018; Ilee et al., 2021). CH$_3$OH has only been detected in three disks (Walsh et al., 2016; van 't Hoff et al., 2018; Booth et al., 2021), and CH$_3$OCH$_3$ was only detected in the Oph IRS 48 disk at the location of the ice/dust trap (Brunken et al., 2022). The simplest organic acid (HCOOH) was detected in the TW Hya protoplanetary disk (Favre et al., 2018). In spatially resolved observations, the COMs seem to be highly concentrated in the inner 50-100au region of the disks (Ilee et al., 2021). In contrast, smaller hydrocarbons and nitriles are more widely distributed.

## Theories and models of chemistry in protoplanetary disks

### Types of chemical reactions in protoplanetary disks

Even in the dense region of protoplanetary disks, the number density is around $10^{12-13}$ cm$^{-3}$, still orders of magnitude lower than that of planetary atmospheres. Furthermore, most region of the disk is cold (<100 K). These low-density and low-temperature environments lead to five important characteristics of chemistry in the bulk regions of protoplanetary disks (Oberg and Bergin, 2021):¨ (1) *dominance of two-body reactions:* The average rates of two molecules colliding together are low, and therefore the reactions are almost exclusively two-body reactions, three-body reactions are extremely rare except for the innermost region (e.g., Kamp et al., 2017). (2) *time-dependent chemistry:* Many chemical reactions take a much longer time than the disk lifetime to reach equilibrium. Therefore, chemical compositions depend on the initial conditions as well as the disk age, deviating significantly from thermochemical equilibrium. (3) *the importance of ions:* Ion reactions are generally much faster than neutral-neutral reactions. As a result, the ionization sources (UV, X-ray, and cosmic-ray ionization) are driving forces of chemistry in protoplanetary disks. (4) *Difficulty of exothermic reactions:* Due to the low temperature, atoms and molecules generally lack sufficient kinetic energy to overcome substantial reaction barriers. Exothermic reactions (reactions that generate energy) are more unlikely and faster. (5) *Grain surface chemistry:* Molecules freeze out onto grain at low temperatures, and ice on grain surface can remain chemically active. Surface chemistry is an important channel for making molecules. These new molecules formed on grain surface can go back into gas-phase through processes such as sublimation.

The gas-phase chemistry is classified into different types, based on the formation, destruction, and rearrangement of chemical bonds (e.g., Herbst and Leung, 1989; Herbst and van Dishoeck, 2009).

Formation of bonds: In two-body reactions ($A+B \rightarrow C$), new bonds can form when photons or electrons carry away the bond-formation energy, which is called *radiative association* and *associative detachments*, respectively.

Destruction of bonds: To break a bond, extra energy needs to be inserted into the molecule first. This extra energy can be obtained by absorption of photons (*photodissociation*), collision with electrons (*dissociative recombination*), and collision with other molecules in shocks. Typical covalent bonds have strengths above 5ev, and therefore UV or X-ray are required to destroy bonds. Besides photon energy, cosmic-ray can also be an important source of energy to destroy bonds.

Rearrangement of bonds: There are reactions that rearrange chemical bonds to form new molecules. The process can be generalized as $AB+C \rightarrow A+BC$. The reactants can be ions, radicals, neutral molecules, and atoms. In most of the disk region, ion-molecule reactions are much faster than neutral-neutral reactions, as ions can cause induced dipole on neutral molecules. Therefore, UV, X-ray, and cosmic-ray are the main producers of ions in disks and therefore are important drivers of chemistry in planet-forming disks. Ion-molecule reactions are central in the chemistry of protoplanetary disks. Most of neutral-neutral reactions only become important in the high-temperature region (T>400K). Nevertheless, some neutral-neutral reactions may be rapid even at low temperatures, especially reactions involve small radicals (e.g., $C_2H$) (e.g., Herbst and Woon, 1997; Smith et al., 2004).

Grain surface reactions: Besides gas-phase reactions, reactions on grain surfaces can play an important role in the formation of molecules, especially in the formation of more complex molecules (e.g., Hasegawa et al., 1992; Garrod et al., 2008). Molecules freeze out onto grain surfaces, building up icy mantles. Light atoms, like H, can migrate on grain surfaces until they find a reactant on the grain surface to form a new molecule. The grains act as a third body to absorb bond formation energy and thus facilitate the formation of new bonds. This type of H addition process is called hydrogenation reaction, which is believed to play an important role in the formation of $H_2$ (e.g., Cazaux and Tielens, 2004) and how water is formed in the cold ISM (e.g., van Dishoeck et al., 2013). Grain surface reactions are also important ways to make complex molecules, while the detailed pathways are still highly uncertain. The molecules formed on the grain surface can be released back into the gas phase, due to thermal desorption, photodesorption, electron-stimulated desorption, and other energetic events.

Chemical modeling of protoplanetary disks

To model the chemistry in protoplanetary disks, we need to know what types of chemical reactions happen (the choice of chemical networks), the local physical conditions (e.g., density, temperature, radiation field, available grain surface area), and the initial compositions of the materials as the chemistry is time-dependent. Furthermore, if the timescales of dynamical processes, such as diffusion, turbulence mixing, and transport of materials, are shorter or comparable to the chemical timescales, these processes should be added as additional source and sink terms.

The chemical reaction rates for chemistry in protoplanetary disks can be found in two main databases: the UMIST Database and the KIDA (McElroy et al., 2013; Wakelam et al., 2015). These databases collect comprehensive reaction rates for gas-phase chemical reactions, including two-body reactions, photodissociation and photoionization, cosmic-ray ionization, and cosmic-ray induced photodissociation and ionization. Some rates are from lab experiments, and many are from theoretical calculations. The rates are temperature dependent but rates at low temperature regions are often missing. Grain-surface reaction rates are still highly incomplete, and the networks often stop at the simplest complex organics (e.g., $CH_3OH$). Commonly used grain-surface reaction networks include the Hasegawa et al. (1992) and the Ohio State University (OSU) network (Garrod et al., 2008). Depending on the environment conditions, the chemical networks are sometimes simplified to accelerate the computational speed (e.g., van 't Hoff et al., 2017; Krijt et al., 2020) or sometimes additional networks are added to account for three-body and high-temperature reactions (e.g., Kamp et al., 2017; Anderson et al., 2021; Kanwar et al., 2023).

Current chemical models often adopt static density structures for gas and dust in protoplanetary disks. The stellar radiation is the dominant energy input for the thermal structure and radiation field in disks. The stellar photospheric spectra are often adopted from theoretical models, and UV and Xray spectra are taken from observations or theoretical models (e.g., Getman et al., 2005; Yang et al., 2012; Dionatos et al., 2019). Once a physical disk structure is set up, stellar spectra over the whole wavelength range (from X-ray to mm wavelength) are used as inputs to compute photon propagation and then determine the radiation field inside the disk (e.g., Woitke et al., 2009; Walsh et al., 2010; Bruderer et al., 2012; Du and Bergin, 2014; Cleeves et al., 2015). Cosmic-ray ionization rate inside the disk usually uses a simple e-fold column density of 96g cm$^{-2}$ from the ISM value of $10^{-17}$ s$^{-1}$ or some fixed values (Umebayashi and Nakano, 1981). The dust temperature structures are usually based on radiative transfer calculation of heating and cooling balance between stellar light and thermal emission of dust grains (Bjorkman and Wood, 2001; Dullemond and Dominik, 2004; Pinte et al., 2006; Bruderer et al., 2012; Du and Bergin, 2014). In the disk atmospheres, the dust and gas no longer have enough collisions to be thermally coupled, and the gas temperature can become much higher than the dust temperature (e.g., Kamp and Dullemond, 2004; Najita et al., 2011; Bruderer et al., 2012). Some models calculate the gas temperature self-consistently based on the heating and cooling processes of gas, while others adopt a semi-analytical formula to correct the temperature differences between gas and dust. The initial conditions are usually set as atomic abundances based on molecular clouds, or atomic ratio and simple CO and $H_2O$ abundances. Readers can find a more comprehensive review of chemical modelings of protoplanetary disks in Henning and Semenov (2013).

Most chemical models run time-dependent chemistry for 1 to a few Myrs and then use the abundance structure to compare with observations. The abundance and thermal structures are usually used as inputs for radiative transfer models to generate simulated observations. The line excitation is assumed to be at local-thermal-equilibrium (LTE) or is calculated to archive collisional balances at the local density and temperature conditions (e.g., Brinch and Hogerheijde, 2010). The line model images from radiative transfer calculations are then convolved with the proper spatial and spectral resolutions to compare with observations.

## Current insight into chemistry in protoplanetary disks snowlines

Snowlines are where molecules freeze out in the disk. Due to radial and vertical temperature gradients in disks, different molecules have snowlines at different distances from the central star. It is believed that the snowlines of the abundant volatile molecules, such as water, $CO_2$, and CO, play an important role in planet formation (e.g., Hayashi, 1981). In particular, the water snowline has long been used to explain the dichotomy of the terrestrial and giant planets in the solar system, i.e., the giant planets formed outside the water snowline because water ices provide extra solid masses and subsequently accelerated the growth of planetary cores (e.g., Stevenson and Lunine, 1988; Pollack et al., 1996; Kennedy and Kenyon, 2008).

There are several arguments for the importance of snowlines in planet formation: (1) *Solid surface density is enhanced beyond snowlines*. Beyond a particular snowline, gas phase molecules freeze out and become ices, which can enhance the surface density of by a factor of 1-4 (depending on the assumed rock-ice ratio in modeling). (2) *snowlines facilitate the formation of planetesimals*. One line of argument is the particles' stickiness and fragmentation thresholds change across different snowlines as their icy composition changes (e.g., Gundlach and Blum, 2015). As a result, the maximum grain sizes increase when the thresholds become higher, or grains become smaller when the thresholds become lower. For example, the fragmentation threshold of bare silicates is expected to be lower than that of water ice grains. When grains cross the water snowline, they become smaller due to the loss of icy mantle and the change of fragmentation threshold. Smaller grains drift inwards more slowly compared to the larger ones, making a local "traffic jam" just inside the snowline (e.g., Banzatti et al., 2015; Pinilla et al., 2017). Also, the ice evaporation and outward diffusion of water increases the abundance of icy pebbles outside the water snowline (Stevenson and Lunine, 1988; Cuzzi and Zahnle, 2004; Ciesla and Cuzzi, 2006; Ros and Johansen, 2013). The enhanced local dust-to-gas ratio around the water snowline may trigger planetesimal formation via streaming instability (e.g., Drazkowska and Alibert, 2017). (3) *snowlines set the boundary of disk regions with different elemental ratios in gas or solids* As different molecules have different snowlines, the partition of C and O elements in the solid and gas phases depends on the distance from the central star. Therefore, a gas giant that accretes its envelop materials between two snowlines would carry a birthmark in their C/O, C/H, and O/H ratios in their atmospheres (Oberg et al., 2011). Please see¨ more discussion on this in the Section on Elemental ratios.

Despite their great importance in planet formation theories, snowlines are challenging to be directly observed. Most of the snowlines like $H_2O$ and $CO_2$ are inside 10au for typical disks, where the dust emission becomes highly optically thick and therefore the line emission is only from the surface layer. The surface layer is expected to be hotter than the mid-plane, unless the disk mid-plane is heated up by very high accretion heating. Several studies measured the water snowline at the disk surface layer, by constraining where the column density of water vapor drops promptly (e.g., Zhang et al., 2013; Blevins et al., 2016). These results suggested the surface water snowlines are between 3-11au. Still, we do not have direct constraints on their mid-plane water snowline. If a disk is ongoing an accretion outburst that dramatically increases its luminosity and subsequently warm up the disk, the water snowline can be moved outwards to 10-100au, and becomes much easier to detect. Recently, HDO and $H^{18}_2O$ lines have been imaged in the V883 Ori outburst disk, suggesting a water snowline at 80au (Tobin et al., 2023). There are also ideas of indirect tracers of the mid-plane water snowlines. One possibility is to use a molecule that is chemically anti-correlated with water vapor abundance. $HCO^+$ has been proposed to be such as a tracer (van 't Hoff et al., 2018; Leemker et al., 2021). Another idea is to look for a sharp change in the spectral slope of continuum emission, as a result of dust size changes around the water snowline (Banzatti et al., 2015; Cieza et al., 2016). But more tests are needed to evaluate the robustness of these methods.

Compared to water and $CO_2$, the mid-plane CO snowline is expected to be at a much larger distance, ~20-40au for disks around solar-like stars and 60-100au for disks around 2-3 $M_\odot$ stars. The abundant $^{12}CO$, $^{13}CO$, or sometimes even $C^{18}O$ become optically thick inside the mid-plane CO snowline, and therefore more rare CO isotopologues are needed to probe the mid-plane CO snowlines. Spatially resolved $^{13}C^{18}O$ line emission has been used to constrain the mid-plane CO snowlines in three disks, founding CO snowline between 20-100au (Zhang et al., 2017; Loomis et al., 2020; Zhang et al., 2021). However, observing the faint CO isotopologue lines is very time-consuming, and therefore, the number of disks with direct CO snowline constraints is still small. Alternatively, $N_2H^+$ can be used as an indirect tracer of the mid-plane CO snowline, because this ion only becomes abundant when the gas-phase CO abundance is low. Spatially resolved $N_2H^+$ line emission often shows an inner cavity, and the radius of the inner cavity is inferred as the mid-plane CO snowline (Qi et al., 2013, 2015, 2019). However, thermo-chemical models show that the cavity radius of $N_2H^+$ is probably an upper limit of the mid-plane CO snowline, and the offset between the two depends on the $CO/N_2$ abundance ratio and the detailed temperature structure in the disk (e.g., van 't Hoff et al., 2017).

## Abundances of carbon, oxygen, and nitrogen carriers

Carbon, oxygen, and nitrogen are the three most abundant elements beyond H and He. The molecules and ions with these elements are the backbones of chemistry in planet-forming disks, and therefore understanding the abundances and distributions of these carriers is an essential part of our understanding of chemistry in planet formation (e.g., Pontoppidan et al., 2014; Krijt et al., 2023; Oberg et al., 2023).

The initial elemental abundances of a planetary forming disk are assumed to be the same as that of its central star, considering the star and its disk are formed from the same molecular cloud. Measuring the CNO elemental abundances of individual young stars is challenging because the active accretion fills out the absorption lines in stellar photospheres. Due to this challenge, the total CNO elemental abundances are often assumed to be the solar abundances. The fractions of different CNO carriers at the birth time of protoplanetary disks are estimated based on ice compositions seen in protostellar envelopes and in solar system comets (e.g., Oberg et al., 2011; Mumma and Charnley, 2011; Boogert et al., 2015). Below we discuss current understanding of the main CNO carriers in protoplanetary disks.

Carbon: Nearly half of the carbon atoms are thought to be in a refractory format, based on observations of carbon depletion in diffuse interstellar medium (Mishra and Li, 2015). The rest of the carbon atoms are in volatile formats (in gas or ice), such as CO, $CO_2$, and $CH_4$ (e.g., Pontoppidan et al., 2008; Oberg et al., 2011; Boogert et al., 2015). CO is the main carbon carrier in a volatile¨ format. Indeed, CO is the most widely detected molecule in protoplanetary disks.

For many years, the CO abundance in disks has been assumed to be the canonical interstellar medium ratio of CO-to-$H_2$ ratio of $10^{-4}$. However, recent ALMA surveys of 1-3Myr-old protoplanetary disks showed that the $^{13}CO$ and $C^{18}O$ line fluxes are much lower than expected (e.g., Favre et al., 2013; Ansdell et al., 2016; Long et al., 2017), even after the correction of photodissociation and freeze-out (Miotello et al., 2017). One possibility is that the gas in disks dissipates much faster than previously assumed and therefore gas-to-dust mass ratios in these disks are low. However, observations showed that many of these disks still have relatively high accretion rates onto their central stars (~$10^{-9}$ $M_\odot$/year). A low gas mass disk would run out of gas extremely fast (0.1Myr), which is in conflict with the age of 1-3Myr of these disks (Manara et al., 2016, 2020). More direct evidence is three protoplanetary disks have independent gas mass constraints from HD (1-0) line fluxes (Bergin et al., 2013; McClure et al., 2016). Their $C^{18}O$ line flux suggests that the CO-to-$H_2$ ratios in these three disks are 1-2 orders of magnitude lower than the canonic ISM ratio (Favre et al., 2013; Schwarz et al., 2016; Calahan et al., 2021; Trapman et al., 2022). Zhang et al. (2020); Bergner et al. (2020) studied CO abundances of several youngest disks (<1Myr), showing that the CO-to-$H_2$ ratio starts with the ISM ratio at the youngest disks, but rapidly decrease with ages, with a timescale of ~1Myr.

To explain the low CO gas abundances in 1-3 Myr-old disks, two broad types of mechanisms have been proposed. The first type is chemical processes. The idea is that once CO gas freezes out onto grain surfaces, surface reactions can process CO into other molecules like $CO_2$, $CH_4$, and $CH_3OH$ (e.g., Bergin et al., 2015; Schwarz et al., 2018; Bosman et al., 2018). This chemical process require a cosmic ray ionization rate of $10^{-17}$ s$^{-1}$ or higher to reduce the CO abundance by one order of magnitude within 1Myr. However, it is still unclear if such a high cosmic ionization rate exists in disks (e.g., Cleeves et al., 2013, 2015; Seifert et al., 2021; Aikawa et al., 2021). The second type of mechanism proposed is dust growth. As dust grains grow into larger sizes, they settle onto the mid-plane of the disk. As a result, CO ices are carried with icy grains onto the mid-plane and gradually reduces the CO abundance in the disk atmosphere. The problem is still the efficiency, as current dust evolution models only predict a factor of few depletion in 1Myr (e.g., Xu et al., 2017; Krijt et al., 2018). However, the two types of mechanisms may work together. Recent simulations including both chemical processes and dust growth showed the CO abundance can be depleted by 2 orders of magnitude in 1Myr timescale (Krijt et al., 2020).

The $CO_2$ is another main carbon-carrier beyond CO. $CO_2$ gas has been detected via its rovibrational Q-branch around 15$\mu$m in ~10 protoplanetary disks by Spitzer (Pontoppidan et al., 2010; Carr and Najita, 2011). In Spitzer observations, the detection rate of $CO_2$ was much lower than $H_2O$, HCN, and $C_2H_2$ in protoplanetary disks. The $CO_2$ gas abundances in disks are still highly uncertain, as the 15$\mu$m feature is dominated by optically thick emission (Bosman et al., 2017). The inferred $CO_2$-to-$H_2$ abundances are on the order of $10^{-9}-10^{-7}$, much lower than the expected ISM value of $10^{-5}$ (Bosman et al., 2017). Recently, $^{13}CO_2$ was detected in the JWST/MIRI spectra of the GW Lup protoplanetary disk, which suggested that the $CO_2$ abundance is orders of magnitude higher than previously derived from the Spitzer data (Grant et al., 2023). In the next few years, JWST observations will provide more accurate constraints on the $CO_2$ abundances in protoplanetary disks.

$CH_4$ is expected to be another main carriers of carbon in protoplanetary disks. However, it has only been detected in near-IR absorption features in the GV Tau N protoplanetary disk (Gibb and Horne, 2013). $CH_4$ has vibrational bands at 7.5$\mu$m, where the spectral resolution of Spitzer/IRS was only 100 and did not provide sensitive constraints. JWST/MIRI (with R~ 3000), will provide one order of magnitude better constraints on the $CH_4$ abundances in a large number of protoplanetary disks.

Oxygen: Based on extinction and absorption feature studies in the ISM, about 25% of the atomic oxygen is locked in silicates, 40% is in an unknown format, and the rest of 35% in volatiles, mainly $H_2O$, $CO_2$, and CO (Whittet, 2010; Boogert et al., 2015). $H_2O$, CO, and $CO_2$ take ~ 10%, 25% and 10% of the oxygen budget, respectively.

Similar to the CO depletion seen in disks, the observations of far-IR water in disks suggested a significant depletion of $H_2O$ in the atmosphere of the outer disk region (50au and beyond) (Hogerheijde et al., 2011; Du et al., 2017). Beyond the mid-plane water snowline, a low abundance of water vapor exists as UV radiation can photodescorb water from dust grain surfaces (e.g., Walsh et al., 2010). However, deep searches of cold water vapor in protoplanetary disks by the Herschel space telescope have very low detection rates (e.g., Bergin et al., 2010; Hogerheijde et al., 2011). Compared with predictions from thermo-chemical models, the results suggested that the water vapor abundance is 1-2 orders of magnitude lower than the models with ISM level O/H elemental ratio (Du et al., 2017). The depletion of water vapor is also consistent with the expectations of dust growth and sequestering water ice onto the mid-plane (Krijt et al., 2016).

In the inner disk region (< 5au), the abundances of carbon and oxygen carriers are still highly uncertain. The main tracers of the inner warm disk regions are mid-IR molecular lines, but these lines are emitted from the surface disk layer, probably only represent a few percent of the total column density (Woitke et al., 2018; Bosman et al., 2022). Current dust evolution models suggest that large amount of icy grains drift into the inner disk region, and the inner disk can be enriched in carbon and oxygen after ices evaporate (e.g., Booth et al., 2017; Krijt et al., 2018; Booth and Ilee, 2019; Krijt et al., 2020). Trends of higher water column density and lower C/O ratios have been seen in more compact disks and less massive disks (Banzatti et al., 2020). This is consistent with simulations of dust evolution in disks. Upcoming JWST observations can test this scenario in a large number of disks.

Nitrogen: In protoplanetary disks, the majority of nitrogen is expected to be in the format of molecular nitrogen ($N_2$), based on $N_2/NH_3$ ratio measured in star-forming molecular clouds (Womack et al., 1992). Unfortunately, $N_2$ is hard to be directly detected as it does not have a permanent dipole moment and therefore lacks of strong lines. Therefore, tracer species are needed to understand the nitrogen abundance, with the assistance of chemical models. $NH_3$ and HCN are the most abundant N-carriers detected in comets (Mumma and Charnley, 2011) and therefore natural candidates of tracer species. $NH_3$ has only been detected in one protoplanetary disk (Salinas et al., 2016), and upper limits of $NH_3$/H were reported for a few disks (Pontoppidan et al., 2019). In contrast, HCN is widely detected in nearby disks, both in its vibrational branch at mid-IR and pure rotational lines at millimeter wavelengths (e.g., Salyk et al., 2011; Huang et al., 2017; Guzman et al., 2021). Therefore, HCN and its isotopologues have been used to constrain the N/H elemental ratio in protoplanetary disks (e.g., Cleeves et al., 2018). CN is already readily detectable in disks (e.g., Dutrey et al., 1997; Oberg et al., 2010; Cazzoletti et al., 2018). $N_2H^+$ is another commonly detected nitrogen carrier in protoplanetary disks (e.g., Qi et al., 2013, 2019; Anderson et al., 2019), but its abundance depends on both the $N_2$ abundance and the cosmic-ray ionization rates in disks. The usage of different N-carriers to constrain Nitrogen abundance is still relatively new and more tests are needed.

In general, $N_2$ is expected to stay in gas phase in the bulk region of the disk and less subjected to depletion in gas phase, due to its low condensation temperature (~19K). Indeed, the elemental ratios seen in solar system bodies showed orders of magnitude depletion of N, suggesting that in the solar nebula most of N likely stayed in gas phase (e.g., Bergin et al., 2015). Chemical models of CO isotopologues and HCN observations of the IM Lup protoplanetary disk shows that its C/H ratio in gas is depleted by a factor of 10-20, but N/H ratio is still close to the ISM ratio of $10^{-5}$ (Cleeves et al., 2018). Combined $N_2H^+$ and CO isotopologue studies in two old Upper Sco disks also suggested that their CO gas abundances are heavily depleted but N/H is consistent with the ISM ratio (Anderson et al., 2019).

## Complex Organic Molecules

Understanding the abundances and distributions of complex organic molecules (COMs) in protoplanetary disks is of great importance because the COMs in the Solar Nebular were potential seeds of life on Earth. The early Earth may have obtained most of its volatile materials at its surface from meteorites and comets (e.g., Anders, 1989; Altwegg et al., 1999). Both in-situ and remote observations have revealed that comets and carbonaceous meteorites are rich in COMs. About a quarter of comet mass is in the format of complex organic refractory materials (dominated by carbon), 9% in small carbonaceous molecules, and a small contribution from simplest organics in water ice (e.g., Greenberg, 1998). Most of the organics found in carbonaceous meteorites are macro-molecules (Cronin and Chang, 1993). COMs have been widely seen in star-forming regions, both in cold gas in large scales, and in the inner envelop region close to forming protoplanetary disks (see recent reviews by Jørgensen et al., 2020). However, it is still unclear what the COMs abundances and distributions are in protoplanetary disks, and how they may depend on the stellar masses and/or the stellar environments.

The observations of COMs in protoplanetary disks have been challenging, because of three reasons. The first reason is that the temperature of the bulk region of the disk is below 100K and therefore most of the COMs exist as ices and cannot be probed by molecular emission line observations. The second reason is that large molecules have more degrees of freedom and more possible transitions, which dilutes the strength of individual transitions and makes their detections more challenging than the simple molecules. Finally, the sizes of protoplanetary disks are small, up to a few hundred of AUs, which is much smaller compared to the sizes of protostellar envelopes (a few thousand AUs) and the dark clouds in the ISM.

Given these challenges, the number of COM detections in protoplanetary disks is still small and mostly limited to the most luminous and massive disks. In the outer disk region, the most commonly detected COMs are $HC_3N$, $CH_3CN$, and $c-C_3H_2$ (e.g., Chapillon et al., 2012; Oberg et al., 2015; Bergin et al., 2016; Bergner et al., 2018; Ilee et al., 2021), which have been detected in ~20 disks. Spatially resolved observations showed that line emissions of these organics tend to concentrate at the inner 50-100au region, more compact compared to the emitting areas of simple molecules, such as CO, CS, and $H_2CO$. In contrast, $CH_3OH$ has only been detected two disks between (110Myr), one in a T Tauri disk (Walsh et al., 2016) and the other one in a Herbig disk (Booth et al., 2021). Compared to the older disks, the young embedded disks (<1Myr) are warmer and therefore there are better chances to detect COMs (Podio et al., 2020). In particular, multiple COMs lines, including $CH_3OH$, $CH_3CHO$, $CH_3OCHO$, and $CH_3COCH_3$, were detected in the V883 Ori disk, an FU Ori type source that has an on-going accretion outburst with an enhanced luminosity of $200L_\odot$ (van 't Hoff et al., 2018; Lee et al., 2019). So far, most of the COMs detections of protoplanetary disks have been at millimeter wavelengths, tracing line emission outside of 10au.

However, the recent JWST/MIRI detections of multiple COMs in a M-dwarf disk, including $C_4H_2$, $C_6H_6$, and $^{13}C^{12}CH_2$ (Tabone et al., 2023). This may open a new window to study COMs in the inner few AU region of disk region.

The origin of COMs in protoplanetary disks is still under debate. One fundamental debate centered on how much the COM compositions are inherited from the interstellar and prestellar materials and how much the compositions are altered by chemical reactions inside the disk. It is generally thought a significant amount of materials may be inherited from the ISM and prestellar, as suggested by the similarity of relative compositions in comets and protostellar inventories (e.g., Mumma and Charnley, 2011; Drozdovskaya et al., 2019). The COM ratios found in the outburst source V883 Ori disk also broadly agree with cometary values, consistent with the picture of the observed COMs are sublimated from inherited organic ices from the protostellar stage. On the other hand, chemical models suggest that there may be significant CO freeze-out onto grain surfaces, and these carbons can be processed into $CH_3OH$ or more complex organics. Besides grain surface reactions, photochemical-dominated gas-phase chemistry may develop at the later stage of disk evolution, as dust grains grow, settle, and drift. This high gas-dust-ratio and C-rich environment allows for deeper UV penetration and therefore gas-phase formation of organic molecules (Calahan et al., 2023).

## Elemental ratios: link from disk compositions to planetary compositions

One of the ultimate goals of planet formation studies is to understand how the initial conditions and planet formation processes determine planetary compositions. There are significant challenges to this mighty goal. Exoplanetary systems are known to be incredibly diverse, even within a narrow stellar mass range. This diversity could be the result of different initial conditions, formation history, and/or stochastic processes in planet formation. The hope is that some observable features of a planet are unique results of its initial conditions or formation history, and therefore we can use the observed characteristics to inform its formation history and deduct other important but unobservable properties of the planet, such as its interior structure and compositions and its surface conditions.

## Link between disk compositions and exoplanetary atmospheric compositions

So far, the elemental ratios of carbon and oxygen elements (C/O, C/H, and O/H) have been the most widely explored tracers to connect the atmospheric compositions of hot-Jupiters and their formation/migration history (e.g., Oberg et al., 2011; Madhusudhan, 2012; Mordasini et al., 2016). Volatile molecules (e.g., $H_2O$, $CO_2$, CO, $CH_4$) are the most detectable molecules in exoplanets' atmospheres (see review by Madhusudhan, 2019). The C/O elemental ratio can be measured in hydrogen-dominated atmospheres, as the C/O ratio critically influences the relative abundances of dominant species, like $H_2O$, $CO_2$, CO, $CH_4$ (Madhusudhan, 2012). On the other hand, molecules of Carbon, Nitrogen, and Oxygen (CNO) elements account for the majority of the mass of ices and gaseous species (beyond H and He) in protoplanetary disks (e.g., Salyk et al., 2013; Pontoppidan et al., 2014). Given their chemical importance and detectability of carbon and oxygen, C/O elemental ratio has been proposed as a key connective tissue between atmospheric compositions of planets and their formation/migration history (e.g., Oberg et al., 2011; Madhusudhan, 2012; Mordasini̇ et al., 2016; Cridland et al., 2020).

The key idea is that the C/O elemental ratio in the atmosphere of a gas giant is a birthmark from the planet's forming region in the natal disk. The simplest argument is that the main ice species $H_2O$, $CO_2$, CO, have different condensation temperatures, and therefore their snowlines are located at different distances from the central star. As a result, the elemental ratio of C/O in the gas-phase (solid-phase) of disk materials increases (decreases) like a multi-step function between the snowlines (see Figure 3). If a gas giant planet accretes the bulk of its atmosphere masses within a region between two snowlines and there is no significant mixing between the planetary core and atmospheric materials, the C/O ratio of its atmosphere can tell its forming region (Oberg̈ et al., 2011). For example, a giant planet with C/O ≥1 can only form in the region outside the CO snowline. However, this simple scenario did not consider the effects of time-dependent chemistry in disks, the transport of solid materials, and the alteration of volatile compositions during planet formation processes.

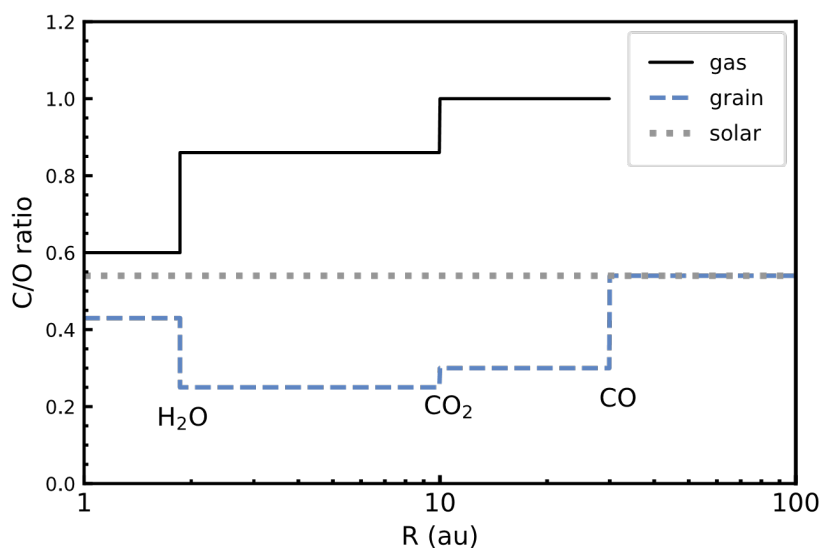

Figure 3: How the C/O elemental ratio in the gas and grains change at different radii in a typical protoplanetary disk around a solar-like star. The snowlines of $H_2O$, $CO_2$, and CO are boundaries of different elemental ratios. Figure is adapted from Oberg et al. (2011).

Since then, many works have explored the impacts of chemistry and formation processes on the C/O ratio in the planetary atmosphere, using different levels of complexity and considering various mechanisms of planet formation processes (e.g., Madhusudhan et al., 2014; Oberg and̈

Bergin, 2016; Mordasini et al., 2016; Cridland et al., 2016; Booth et al., 2017; Booth and Ilee, 2019; Cridland et al., 2020; Notsu et al., 2020). However, when different assumptions and processes are considered, the resulting C/O are C/H ratios are not always the same. For example, whether planetary atmospheric composition is dominated by the composition of the accreted gas can lead to distinctive predictions on the C/O ratio in planetary atmospheres. Oberg et al. (2011) and Madhusudhan et al. (2014) predicted that C/O≥1 can be achieved if a planet forms beyond the CO snowline, while other studies showed that if planetesimal accretion dominates the volatile and metal enrichment in planetary atmospheres, all of the models predict subsolar C/O ratio (Thiabaud et al., 2015; Morbidelli et al., 2016). Another question is whether the disk materials move across the disk. In the static disk models, the absolute ratios of C/H and O/H in the disk gas are always below the stellar ratio across the whole disk, because some C and O are locked in refractory materials and the rest of the C and O are further locked in ices at larger distance from the central star. In contrast, in models with pebble drift, a large amount of icy pebbles drift into the inner disk region. When icy pebbles cross a snowline, the corresponding ice mantle can evaporate and enrich the volatile abundances in the gas phase. As a result, the C/H or O/H ratios in the disk gas can become super-stellar (Oberg and Bergin, 2016; Booth et al., 2017; Booth and Ilee, 2019; Estrada and Cuzzi, 2022; Bitsch and Mah, 2023). Indeed, this super-stellar C/H ratio has been seen in the gas inside the CO snowline of the HD 163296 protoplanetary disk (Zhang et al., 2020). An enhanced cold water components have been seen JWST/MIRI spectra of two compact disks compared with two extend disks, which is consistent with the expectation that more pebble drift occures in the compact disks (Banzatti et al., 2023). These preliminary evidence require a more robust understanding of planet formation processes.

The assumption of the disk compositions is another main source of uncertainty of the C/O ratio as a tracer of planet birth region. The simplest models adopt the same chemical partition of C and O elements as that of the interstellar ices, assuming full heritage (e.g., Oberg et al., 2011), or¨ assume chemical equilibrium based on the solar abundance, partitioning elements into refractory and volatiles at the mid-plane (e.g., Madhusudhan et al., 2014). The next level of complexity adopts gas and ice compositions from kinetic disk chemistry models after evolution time (typically 1Myr) and assumes no further evolution (e.g., Cridland et al., 2019; Turrini et al., 2021). More complex models have started to consider the coupling between dust growth and chemistry, but so far have only considered simplified chemical networks (e.g., Krijt et al., 2020).

Besides the C/O and C/H ratios, recent studies have started to explore constraints from other abundant elements. Oberg and Wordsworth (2019) and Bosman et al. (2019) argued that the enrichment of Nitrogen in Jupiter's atmosphere requires that the core of Jupiter formed exterior to the $N_2$ snowline beyond 30au, because the main carrier Nitrogen carrier $N_2$ only condenses at very low temperature. Crossfield (2023) proposed that C/S and O/S ratios can be constrained in exoplanet atmospheres via $SO_2$ and provide additional constraints to the formation history. Another idea is to add the Si/H ratio in addition to the C/O and C/H ratios to trace the planet formation history, as Si/H provides a baseline estimation of refractory materials accreted during planet formation (Chachan et al., 2023).

## Isotopic ratios: Link between disk compositions and solar system objects

Many clues of the solar system formation can be obtained from isotopic ratios and the elemental ratios between radiative decay parent and daughter elements (e.g., Marty, 2012; Furi and Marty, 2015; Kruijer et al., 2017). Compared to high precision measurements of elemental and isotopic ratio for the solar system objects, the isotopic ratio measurements of protoplanetary disks are limited to the abundant volatiles (H, C, O, N, and S) and have much higher uncertainties.

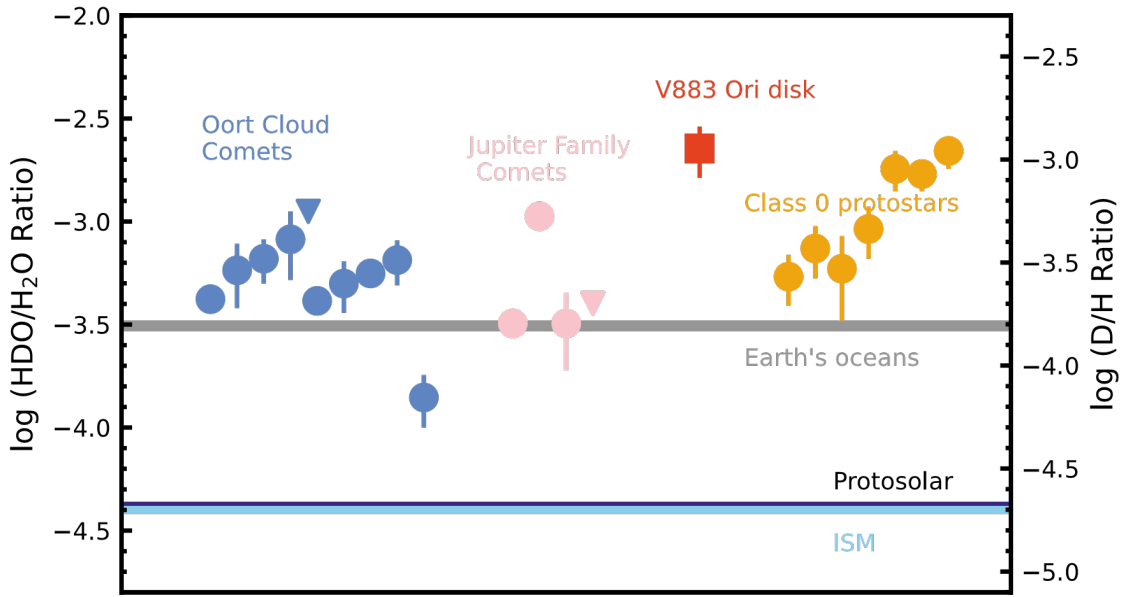

Figure 4: HDO:$H_2O$ ratio in the V883 Ori protoplanetary disk, Class 0 protostars, and different objects in the solar system. Figure is adapted from Tobin et al. (2023).

D/H ratio: The D/H ratio in water has long been used to trace the origin of water on the Earth, by comparing the D/H ratio in Earth's oceans with that in a wide range of solar system bodies, including comets, ancient Martian water, and hydrated minerals in meteorites. The ocean water on Earth on average has a D/H ratio of $1.56\times10^{-4}$, much higher than the D/H elemental ratio of $2.5\times10^{-5}$ measured in the Sun and the D/H ratio of $1.5\times10^{-5}$ within 100 pc of the solar neighborhood (Linsky, 1998). In contrast, the D/H ratio in water at prestellar cores and protostar envelopes, is one to two orders of magnitude higher than the ratio in Earth ocean water, generally between $10^{-3}$–$10^{-2}$ (see review by Ceccarelli et al., 2014).

This enhanced D/H ratio, so-called deuterium fractionation, is widely observed in prestellar and protostar environments. This enhancement is attributed to the result of cold-temperature and ion-driven chemistry (see review by Ceccarelli et al., 2014). As mentioned in Section 3, in cold molecular gas, chemical reactions with ions are much faster than neutral-neutral reactions. $H^+_3$ is the most important ion, which is produced by $H_2$ and H ionized by cosmic ray. It reacts with HD to form $H_2D^+$, i.e., $H^+_3 + HD \leftrightarrow H_2D^+ + H_2 + \Delta E$. In cold gas (≤30K), the ratio of $H_2D^+/H^+_3$ becomes higher than $HD/H_2$, because the ions do not have sufficient energy to overcome the energy barrier ($\Delta E$ ~124K) in the reverse direction of the reaction. When $H^+_3$ and $H_2D^+$ have dissociative recombination with electrons, the H and D atoms are then liberated. The liberated D atoms can then react with O on the grain surface to form HDO. Therefore, cold prestellar cores and protostar envelopes readily produce a high D/H ratio in water ice formed in these environments.

The cold environment can also be provided in the outer region of the protoplanetary disks (e.g., Drouart et al., 1999; Hersant et al., 2001; Yang et al., 2013). However, the cosmic ray ionization rate inside protoplanetary disks may be reduced by one to several orders of magnitude due to stellar and disk winds (Cleeves et al., 2013, 2015). If this is the case, current models predict the ion-drive chemistry in protoplanetary disks cannot produce an enhanced D/H ratio like that of the Earth's ocean water and therefore much of the enhanced D/H ratio must be inherited from the prestellar and protostar stage (Cleeves et al., 2014). However, recent chemical studies of several ion lines in the IM Lup protoplanetary disk suggested that the cosmic ray ionization rate reaches the ISM level between 100-300au region of the disk (Seifert et al., 2021). In general, the cosmic-ray ionization rate in protoplanetary disks is still highly uncertain.

So far, there is only one direct HDO detection in a protoplanetary disk (V883 Ori disk). This disk is undergoing an accretion outburst and therefore has a water snowline at 80au instead of a few au for typical quiescent disks. The measured D/H ratio of $H_2O$ in the V883 Ori disk is consistent with that of protostars (see Figure 4, suggesting significant $H_2O$ in this disk is inherited from the protostar stage (Tobin et al., 2023).

$^{15}N/^{14}N$ ratio: The $^{15}N/^{14}N$ ratio of the Earth's atmosphere clearly differs from those of the protosolar nebula and most of the cometary values, and therefore provide important clues on the formation history of the Earth (Marty, 2012). The $^{15}N/^{14}N$ ratio of the Earth's atmosphere is about $3.678 \times 10^{-3}$, 1.6 times higher than the protosolar value, but about 1.8 times lower than typical ratios measured in comets. Not only the Earth, but all inner solar system objects show enrichment of $^{15}N/^{14}N$ ratio compared to the protosolar nebula value. In protoplanetary disks, the $^{15}N/^{14}N$ ratio has been measured in several disks, based on $HC^{15}N/HCN$ or $C^{15}N/CN$ (Hily-Blant et al., 2017, 2019; Guzman et al., 2017; Booth et al., 2019). The $^{15}N/^{14}N$ ratios measured from HCN are broadly consistent with the ratios in comets, and always are enriched compared to the ratio of the Earth. In one disk where $^{15}N/^{14}N$ ratios were measured from both HCN and CN, the $^{15}N/^{14}N$ from CN is much less enriched in $^{15}N$ compared to HCN (Hily-Blant et al., 2017). In two protoplanetary disks, the radial distribution of $^{15}N/^{14}N$ ratio has been measured from spatially resolved $HC^{15}N$ images. Both cases showed that $^{15}N/^{14}N$ ratio shows a strong radial gradient, with the inner disk region being 2-3 times more enriched in $^{15}N$ (Guzman et al., 2017; ́Hily-Blant et al., 2019).

$^{18}O/^{16}O$, $^{17}O/^{16}O$ ratios: Similar to the isotopes of Hydrogen and Nitrogen, Oxygen isotope ratios ($^{18}O/^{16}O$, $^{17}O/^{16}O$) also show variations among bodies in the solar system, providing critical insights into the formation, evolution, and distribution of materials in the solar nebula (e.g.,

Thiemens, 2006). The Sun is known to be $^{16}$O enriched, compared to the terrestrial planets, asteroids, and chondrules (e.g., McKeegan et al., 2011). A possible mechanism to account for the $^{18}$O and $^{17}$O enrichment in solar system bodies is self-shielding of CO (e.g., Thiemens and Heidenreich, 1983; Clayton, 2002). CO can be only be photodissociated by UV radiation as specific wavelengths, and different CO isotopologues are photodissociated by UV radiation at lightly different wavelength (van Dishoeck and Black, 1988). Thus, when gas is irradiated by UV radiation, the optically depth of the $^{12}$CO quickly becomes optically thick while less abundant CO isotopologues, such as $^{13}$CO, C$^{18}$O, C$^{17}$O in the interior of the gas still experience photodissociation. This process frees more $^{18}$O and $^{17}$O that go into water ice or other condensates. It is still under debate whether the process mainly happen in molecular cloud or protoplanetary disk stages. Compared to the D/H and $^{15}$N/$^{14}$N ratios, the difference of $^{18}$O/$^{16}$O or $^{17}$O/$^{16}$O ratios compared to the Sun is only upto a few percent level. As discussed above, due the large temperature and abundance gradients in protoplanetary disks, the current abundance retrieval methods of protoplanetary disks have not yet had precision like that of solar system bodies.

## Predictions of future directions

The next decade represents tremendous opportunities to advance the study of chemistry in planet-forming disks and our understanding of the chemical origins of extrasolar systems. We highlight a few promising directions here.

- Transport of volatiles in protoplanetary disks: One exciting opportunity is to test whether the volatile reservoirs in the terrestrial planet-forming region are enriched by large amounts of icy pebbles drift into the inner disk regions, as required by current pebble accretion models of planet formation. It is still unknown whether such significant pebble drift into the inner disk region can widely occur in protoplanetary disks. Evidence of volatile enrichment has been seen in a few cases: in the HD 163296 protoplanetary disk, the C/H ratio inside the CO snowline seems to be enhanced to superstellar level (Zhang et al., 2020, 2021); recent JWST/MIRI spectra of four protoplanetary disks show that an extra excess of cold water emissions are seen in two compact disks compared to two large disks (Banzatti et al., 2023). In the outer disk region (>20au), ALMA can measure chemical structures and constrain the spatial distributions of pebble trapping. In the terrestrial region (inner 5au), JWST can trace the emission of water and simple organics from the surface layer. By combining the power of JWST and ALMA, we can trace the volatile budgets across the whole disk, and test the transport of volatiles in a large number of disks.

- Chemical evolution over the disk lifetime: The gas in protoplanetary disks can last for 110Myr. The chemical composition of disk materials may change dramatically over the disk lifetime (e.g., Eistrup et al., 2016), driven by changes in physical conditions (e.g., gas-to-dust mass ratio, UV intensity, temperature), as well as dynamical processes inside the disk (e.g., turbulence and radial/vertical transport of materials). The lack of empirical constraints on volatile evolution over the disk lifetime poses a great challenge to link atmospheric compositions of transiting planets with their formation history. ALMA already showed some evidence of the CO gas abundance decreases at the timescale of 1Myr (Zhang et al., 2020; Bergner et al., 2020). However, the sample is still small, particularly the the number of youngest (<1Myr) and the oldest disks (5-10Myr) with CO studies. Except for CO, the existing ALMA observations of molecular lines are predominantly towards the large massive disks as well as the intermediate age (1-3Myr) disks. A similar problem also existed in the Spitzer mid-IR spectroscopic studies of protoplanetary disks, largely due to the sensitivity limit. JWST/MIRI has orders of magnitude better sensitivity than Spitzer/IRS and therefore can probe fainter old disks. For young embedded disks, JWST provides better spatial resolution to separate the contributions from the disk and envelope regions. Therefore, in the next few years, not only the number of disks with chemistry will increase significantly, but also the age range as well as the stellar type range will dramatically expand. With a more diverse sample, there will be significant progress in the understanding of chemical evolution over the disk lifetime.

- Coupling of physical and chemical models for protoplanetary disks: So far, the modeling of physical and chemical processes in disks has been largely two separate camps. In the physical modeling camp, sophisticated models, such as dust growth, planetesimal formation, MHD simulations of turbulence and disk wind, and planet-disk interactions, include detailed treatment of physical processes, but often consider minimum or no chemistry in the simulations. On the other side, thermo-chemical models of protoplanetary disks compute time-dependent chemistry using large chemical networks and self-consistent calculate heating-cooling processes. However, most of the chemical models are based on static physical structures and fixed dust populations. Growing observations demonstrate that the physical and chemical processes are probably tightly coupled in planet formation. Recent chemical studies have started to include turbulent mixing, disk wind, and dust growth models, but often used a simplified chemical network (e.g., Semenov and Wiebe, 2011; Heinzeller et al., 2011; Furuya and Aikawa, 2014; Price et al., 2020; Krijt et al., 2020; Eistrup et al., 2022; Van Clepper et al., 2022). In the upcoming decade, the coupling the physical and chemical modeling for planet formation is likely to become a new frontier.

- Linking atmospheric compositions of nascent planets and the chemistry in the local disk areas: To trace the chemical origin of planets, it is of great import to compare the atmospheric compositions of nascent planets to the compositions of disk materials around their forming location. Although the number of nascent planets detected in protoplanetary disks is still small, the number may quickly expand in the next decade. Embedded planets can now also be searched by tracing kinematic perturbation in the disk velocity field. With the new JWST imaging power and the upcoming 30m-class ground-based telescope, we expect to discover more nascent planets while their natal disks are still around and follow up with studies of their atmospheric compositions. Since their natal gaseous disk is still around, we can also probe the chemical compositions around the location of the nascent planet. This type of study has been started for planets inside the PDS 70 protoplanetary disk (e.g., Cridland et al., 2023).

Scanning Disk Rings and Winds in CO at 0.01-10 au: A High-resolution M-band Spectroscopy Survey with IRTF-iSHELL. Astro. J. *163*(4), 174.

Banzatti, A., I. Pascucci, A. D. Bosman, P. Pinilla, C. Salyk, G. J. Herczeg, K. M. Pontoppidan, I. Vazquez, A. Watkins, S. Krijt, N. Hendler, and F. Long (2020, November). Hints for Icy Pebble Migration Feeding an Oxygen-rich Chemistry in the Inner Planet-forming Region of Disks. Ap. J. *903*(2), 124.

Banzatti, A., P. Pinilla, L. Ricci, K. M. Pontoppidan, T. Birnstiel, and F. Ciesla (2015, December). Direct Imaging of the Water Snow Line at the Time of Planet Formation using Two ALMA Continuum Bands. Ap. J. *815*, L15.

Banzatti, A. and K. M. Pontoppidan (2015, August). An Empirical Sequence of Disk Gap Opening Revealed by Rovibrational CO. Ap. J. *809*(2), 167.

Banzatti, A., K. M. Pontoppidan, J. Pere Ch´avez, C. Salyk, L. Diehl, S. Bruderer, G. J. Herczeg,´ A. Carmona, I. Pascucci, S. Brittain, S. Jensen, S. Grant, E. F. van Dishoeck, I. Kamp, A. D. Bosman, K. I. Oberg, G. A. Blake, M. R. Meyer, E. Gaidos, A. Boogert, J. T. Rayner, and¨ C. Wheeler (2023, February). The Kinematics and Excitation of Infrared Water Vapor Emission from Planet-forming Disks: Results from Spectrally Resolved Surveys and Guidelines for JWST Spectra. Astro. J. *165*(2), 72.

Barenfeld, S. A., J. M. Carpenter, L. Ricci, and A. Isella (2016, August). ALMA Observations of Circumstellar Disks in the Upper Scorpius OB Association. Ap. J. *827*, 142.

Bergin, E. A., Y. Aikawa, G. A. Blake, and E. F. van Dishoeck (2007, January). The Chemical Evolution of Protoplanetary Disks. In B. Reipurth, D. Jewitt, and K. Keil (Eds.), *Protostars and Planets V*, pp. 751.

Bergin, E. A., G. A. Blake, F. Ciesla, M. M. Hirschmann, and J. Li (2015, July). Tracing the ingredients for a habitable earth from interstellar space through planet formation. *Proceedings of the National Academy of Science 112*, 8965–8970.

Bergin, E. A., L. I. Cleeves, U. Gorti, K. Zhang, G. A. Blake, J. D. Green, S. M. Andrews, N. J. Evans, II, T. Henning, K. Oberg, K. Pontoppidan, C. Qi, C. Salyk, and E. F. van Dishoeck (2013,¨ January). An old disk still capable of forming a planetary system. Nature *493*, 644–646.

Bergin, E. A., F. Du, L. I. Cleeves, G. A. Blake, K. Schwarz, R. Visser, and K. Zhang (2016, November). Hydrocarbon Emission Rings in Protoplanetary Disks Induced by Dust Evolution. Ap. J. *831*, 101.

Bergin, E. A., M. R. Hogerheijde, C. Brinch, J. Fogel, U. A. Yıldız, L. E. Kristensen, E. F. van Dishoeck, T. A. Bell, G. A. Blake, J. Cernicharo, C. Dominik, D. Lis, G. Melnick, D. Neufeld, O. Panic, J. C. Pearson, R. Bachiller, A. Baudry, M. Benedettini, A. O. Benz, P. Bjerkeli, S. Bon-´ temps, J. Braine, S. Bruderer, P. Caselli, C. Codella, F. Daniel, A. M. di Giorgio, S. D. Doty, P. Encrenaz, M. Fich, A.
25

Brittain, S. D., T. Simon, J. R. Najita, and T. W. Rettig (2007, April). Warm Gas in the Inner Disks around Young Intermediate-Mass Stars. Ap. J. *659*(1), 685–704.

Brown, J. M., K. M. Pontoppidan, E. F. v. van Dishoeck, G. J. Herczeg, G. A. Blake, and A. Smette (2013, June). VLT-CRIRES Survey of Rovibrational CO Emission from Protoplanetary Disks. Ap. J. *770*(2), 94.

Bruderer, S., E. F. van Dishoeck, S. D. Doty, and G. J. Herczeg (2012, May). The warm gas atmosphere of the HD 100546 disk seen by Herschel. Evidence of a gas-rich, carbon-poor atmosphere? Astron. & Ap. *541*, A91.

Brunken, N. G. C., A. S. Booth, M. Leemker, P. Nazari, N. van der Marel, and E. F. van Dishoeck (2022, March). A major asymmetric ice trap in a planet-forming disk. III. First detection of dimethyl ether. Astron. & Ap. *659*, A29.

Calahan, J. K., E. Bergin, K. Zhang, R. Teague, I. Cleeves, J. Bergner, G. A. Blake, P. Cazzoletti, V. Guzman, M. R. Hogerheijde, J. Huang, M. Kama, R. Loomis, K. ́Oberg, C. Qi, E. F. van ̈ Dishoeck, J. Terwisscha van Scheltinga, C. Walsh, and D. Wilner (2021, February). The TW Hya Rosetta Stone Project. III. Resolving the Gaseous Thermal Profile of the Disk. Ap. J. *908*(1), 8.

Calahan, J. K., E. A. Bergin, A. D. Bosman, E. A. Rich, S. M. Andrews, J. B. Bergner, L. I. Cleeves, V. V. Guzman, J. Huang, J. D. Ilee, C. J. Law, R. Le Gal, K. I. ́ Oberg, R. Teague, C. Walsh, D. J. ̈ Wilner, and K. Zhang (2023, January). UV-driven chemistry as a signpost of late-stage planet formation. *Nature Astronomy 7*, 49–56.

Carr, J. S. and J. R. Najita (2008, March). Organic Molecules and Water in the Planet Formation Region of Young Circumstellar Disks. *Science 319*, 1504–.

Carr, J. S. and J. R. Najita (2011, June). Organic Molecules and Water in the Inner Disks of T Tauri Stars. Ap. J. *733*(2), 102.

Carr, J. S., J. R. Najita, and C. Salyk (2018, September). Measuring the Water Snow Line in a Protoplanetary Disk. *Research Notes of the American Astronomical Society 2*(3), 169.

Carr, J. S., A. T. Tokunaga, J. Najita, F. H. Shu, and A. E. Glassgold (1993, July). The Inner-Disk and Stellar Properties of the Young Stellar Object WL 16. Ap. J. Letters *411*, L37.

Cazaux, S. and A. G. G. M. Tielens (2004, March). $H_2$ Formation on Grain Surfaces. Ap. J. *604*(1), 222–237.

Cazzoletti, P., E. F. van Dishoeck, R. Visser, S. Facchini, and S. Bruderer (2018, January). CN rings in full protoplanetary disks around young stars as probes of disk structure. Astron. & Ap. *609*, A93.

Ceccarelli, C., P. Caselli, D. Bockelee-Morvan, O. Mousis, S. Pizzarello, F. Robert, and D. Se-́ menov (2014, January). Deuterium Fractionation: The Ariadne's Thread from the Precollapse Phase to
28